\documentclass[10pt,conference]{IEEEtran}
\usepackage{amsmath}
\usepackage{amsfonts}

\usepackage{array}
\usepackage[caption=false,font=footnotesize,labelfont=rm,textfont=rm]{subfig}
\usepackage{textcomp}
\usepackage{stfloats}
\usepackage{url}
\usepackage{verbatim}
\usepackage{graphicx}
\usepackage{cite}

\usepackage{amssymb}
\usepackage{hyperref}
\usepackage{makecell}
\usepackage{bm}

\usepackage{algorithm}
\usepackage{algorithmicx}
\usepackage[noend]{algpseudocode}

\usepackage{multicol}
\usepackage{multirow}

\usepackage[dvipsnames]{xcolor}

\allowdisplaybreaks[4]

\IEEEoverridecommandlockouts

\begin{document}

\title{Adversarial Training: Enhancing Out-of-Distribution Generalization for Learning Wireless Resource Allocation}
		
\author{
	\thanks{This work was supported by the National Natural Science Foundation of China (NSFC) under Grant 62271024.}
	\IEEEauthorblockN{Shengjie Liu and Chenyang Yang}
	\IEEEauthorblockA{Beihang University, Beijing, China\\Email: \{liushengjie, cyyang\}@buaa.edu.cn}
}
	
\maketitle

\thispagestyle{empty}
	
\begin{abstract}
Unsupervised learning has been extensively adopted to train deep neural networks (DNNs) for learning wireless resource allocation. Yet, the performance of DNNs is vulnerable to distribution shifts between training and test data, e.g., wireless channels. In this paper, we propose an offline unsupervised training method to enhance the out-of-distribution (OOD) generalizability of DNNs. Inspired by adversarial training (AT), the method trains DNNs using progressively identified adversarial examples out of the training distribution. To reflect OOD degradation of a DNN in the context of unsupervised learning, we reformulate the optimization problem of AT. The proposed method is evaluated by learning hybrid precoding. Simulation results showcase the enhanced OOD performance of multiple kinds of DNNs with approximately 5\(\sim\)20\% improvement across various channel distributions, even when the samples only from a single distribution (e.g., Rayleigh fading) are used for training.

\end{abstract}

\vspace{-0.5mm}
\section{Introduction}\label{sec:intro}
Deep learning is widely used for optimizing wireless resource allocation, by designing deep neural networks (DNNs) to learn the mapping from environment parameters (say channels) into optimized variables \cite{SCJ}.  For resource allocation problems with closed-form expressions of objective function and constraints, DNNs can be trained in an unsupervised manner. As a consequence, the labels obtained by solving computation-intensive numerical algorithms are no longer required, and constraint satisfaction can be controlled \cite{SCJ, EisenDual}.

The majority of existing works have assumed that training and test data are identically distributed. However, when DNNs are tested on data drawn from unseen distributions, their out-of-distribution (OOD) performance often degrades obviously. In wireless communications, the distribution of environment parameters, especially channels, may vary rapidly due to many factors such as the changed locations of users and scatterers \cite{DGwirelessSurvey}. As a result, the distributions of training and test data may differ. A strong OOD generalizability enables DNNs to perform well across scenarios without further adaptation, thereby reducing the computational demands at wireless edge.

Growing research efforts have attempted to enhance the OOD generalization of deep learning for wireless applications.

A data design method was proposed in \cite{dataAug} to synthesize channel, based on an observation that DNNs trained on channels with proper power delay profiles are generalized well to other distributions when learning channel estimation. A data augmentation method was proposed in \cite{dataAug-CSIfeed} after finding the major differences among channel distributions that are critical for channel feedback. These methods leverage the knowledge of certain tasks and are hard to be applied to other tasks.

It was reported that deep unfolding networks can achieve acceptable performance on unseen distributions when optimizing hybrid precoding\cite{OOD-RISHBF} and power control\cite{OJ-WMMSE-DU}, even without dedicated design for OOD generalizability. It was shown in \cite{OODhypernet-detection} that an unfolding network with the weights dynamically generated by a hyper-network can achieve preferable OOD performance for signal detection. However, deep unfolding methods are problem- and algorithm-specific, which restricts their applicability.

Feature extractor (which is a DNN) was designed to capture domain-invariant feature, which does not change across distributions, from channel input, when learning reconfigurable intelligent surface configuration selection  \cite{IRM-RIS-Bennis} and indoor localization \cite{DomainAT-localization}. The extracted feature was then fed into another DNN to produce the desired output. Nonetheless, the feature extraction approach is not suited for learning complex resource allocation policies, say precoding \cite{ZBCGC}. Meanwhile, samples from multiple distinct distributions are required for training the feature extractor. 

DNNs can be fine-tuned by transfer learning or meta-learning. The adaptation phase is indispensable. Taking hybrid precoding as an example, a DNN was trained by meta-learning and evaluated for OOD generalization without adaptation, whose performance is only about half of that with sufficiently fine-tuning \cite{MLDG-HBF}.
However, adaptations are often infeasible in dynamic wireless environments \cite{DGwirelessSurvey}, where the channel statistics change in tens of seconds \cite{VTC-channel-model}. This is because collecting and training with up-to-date samples are time-consuming and limited by the computing resource at wireless edge.

To achieve good OOD performance, distributionally robust optimization (DRO) has been introduced into machine learning \cite{CuipengSurvey}. In order to be well-generalized across a range of distributions, DRO optimizes the trainable weights of a DNN to perform better on the worst-case distribution where the DNN deteriorates most severely.
Unfortunately, DRO problem is hard to be solved since it involves the optimization of both the weights and the worst-case distribution~\cite{CuipengSurvey}.

Adversarial training (AT), originally proposed to defend against adversarial attacks, has been used as an easy-to-implement alternative to DRO \cite{DRO-Adv-Gao, DRO-Adv-ICML2021}. Specifically, it was proved that the objective of DRO is upper bounded by the objective of AT asymptotically (say Corollary 2  in \cite{DRO-Adv-Gao}, Theorems 1 and 2 in \cite{DRO-Adv-ICML2021}).
Consequently, DNNs trained by AT can also exhibit favorable OOD generalizability \cite{DRO-Adv-ICML2021}.

Both DRO and AT have been investigated only in the context of supervised learning, and can not be applied to unsupervised learning.
The detection and defense for adversarial attacks has also been studied in wireless AI \cite{AdvSurvey1}. However, the viability of using AT to enhance OOD generalization has not been explored for wireless communications.

In this paper, we resort to AT for improving the OOD generalizability of the DNNs that are trained unsupervisedly for optimizing wireless resource allocation. The major contributions are summarized as follows.
(1) To find the worst-case distribution in unsupervised learning, we reformulate DRO and AT problems, enabling them to capture the ``true'' deterioration of DNNs.
(2) We propose an unsupervised adversarial training (UAT) method to solve the reformulated AT problem. The method finds adversarial examples by leveraging the gradient of optimal objective value even though the optimal policy is unknown.
(3) We evaluate the OOD generalization gain from UAT on different channel distributions by training several DNNs only on a single channel distribution, for optimizing hybrid precoding that maximizes sum-rate (SR) under quality of service (QoS) constraints.

The proposed UAT method has three advantages: (1) It does not require specific knowledge to design or augment data. (2) It does not need to modify the architectures of DNNs, in contrast to deep unfolding and domain-invariant feature learning. (3) It does not rely on diverse data, eliminating the need to access samples collected or generated from different distributions, unlike meta-learning and training feature extractor.

\section{Preliminary}\label{sec:pre}
We first recap the DRO and AT in supervised learning.

Denote ${\bf x}$ as the input and ${\bf y}^*$ as the label for supervised learning, where the pair $({\bf x}, {\bf y}^*)$ follows a given distribution (denoted by $\mathcal{D}$) during training. A DNN denoted by $f_{\bf y}({\bf x}; {\bm \theta})$ can produce an output ${\bf y}$ for each ${\bf x}$, where the trainable weight $\bm \theta$ is optimized through a loss function $L({\bf y}, {\bf y}^*)$ to encourage the output ${\bf y}$ to approach the label ${\bf y}^*$. 

To obtain satisfactory OOD performance on unknown distributions, DRO minimizes the expected loss on the worst-case distribution (denoted by $\tilde{\mathcal{D}}$) that incurs the maximal loss, which is formulated as~\cite{CuipengSurvey} \vspace{-2mm}
\begin{equation}\label{eq:DRO}
    \min_{\bm \theta}\max_{\tilde{\mathcal{D}}\in \mathbb{D}} \mathbb{E}_{({\bf x},{\bf y}^*)\sim \tilde{\mathcal{D}}} \left[ L(f_{\bf y}({\bf x}; {\bm \theta}), {\bf y}^*)\right],
\end{equation}
where $\mathbb{D}=\{\mathcal{D}'|W(\mathcal{D}',\mathcal{D})\leq \varepsilon\}$ is the set of distributions around $\mathcal{D}$, $W(\cdot,\cdot)$ denotes the Wasserstein distance between two distributions, and $\varepsilon$ is a threshold.

Although solving the problem in \eqref{eq:DRO} enables the DNN to achieve minimal loss over all distributions in $\mathbb{D}$, which improves its OOD generalizability, the impossibility of sampling from the unsolved $\tilde{\mathcal{D}}$ renders the problem intractable \cite{CuipengSurvey}.

AT methods find adversarial examples and adjust the trainable weights of DNNs with these examples to prevent possible attacks.
Specifically, AT problem minimizes the expected loss over the most maliciously perturbed inputs (denoted by $\tilde{\bf x}$) \cite{PGD}, which is formulated as follows \vspace{-1mm}
\begin{equation}\label{eq:AT}
    \min_{\bm \theta} \mathbb{E}_{({\bf x}, {\bf y}^*)\sim \mathcal{D}} \left[ \max_{\tilde{\bf x}\in B_{\epsilon}({\bf x})} L(f_{\bf y}(\tilde{\bf x}; {\bm \theta}), {\bf y}^*) \right],
\end{equation}
where $B_{\epsilon}({\bf x}) = \{{\bf x}+{\bm \delta} \,| \left\| {\bm \delta} \right\|_2\leq \epsilon \}$ is the set of allowed perturbations around ${\bf x}$, and $\epsilon$ is the perturbation range.

Differing from \eqref{eq:DRO}, the inner optimization variable $\tilde{\bf x}$  in \eqref{eq:AT} is a sample rather than a distribution, which is referred to as the \emph{adversarial example} of ${\bf x}$.
Since the expectation in \eqref{eq:AT} can be approximated by averaging over samples drawn from $\mathcal{D}$, this problem can be solved by alternating the following two steps \cite{PGD}:
(i) {\bf Find $\tilde{\bf x}$:} It can be obtained using gradient ascent methods, guided by the partial derivative of loss function with respect to ${\bf x}$. (ii) {\bf Update ${\bm \theta}$:} It can be optimized using back-propagation, with the loss averaged over all adversarial examples.


\section{DRO and AT in Unsupervised Learning}\label{sec:unsupervisedAT}\vspace{-1mm}
In this section, we formulate DRO and AT for unsupervised learning, and propose a method for solving the AT problem.

\subsection{Problem Reformulation}\vspace{-1mm}
With environment parameter  ${\bf x}$, resource allocation $\bf y$ can be optimized from the following problem \vspace{-1mm}
\begin{equation}\label{prob:whole}
            \min_{\bf y}~F({\bf x}, {\bf y})~~~ {\rm s.t.}~G_i({\bf x}, {\bf y})\leq 0, i=1,\cdots,N_C.
\end{equation}

The optimal solutions for every value of ${\bf x}$ constitute a function: the optimal \emph{policy} ${\bf y}^* \!=\! f_{{\bf y}^*}({\bf x})$.
When DNN ${\bf y} \!=\! f_{\bf y}({\bf x}; {\bm \theta})$ is trained to learn this policy, ${\bf x}$ follows the distribution~$\mathcal{D}$.

By introducing a \emph{multiplier network} (denoted by $f_{\bm \lambda}({\bf x};{\bm \xi})$),  \emph{primal-dual learning} (PDL) \cite{EisenDual,SCJ} can be adopted to train $f_{\bf y}(\cdot\,; {\bm \theta})$ and $f_{\bm \lambda}(\cdot\,;{\bm \xi})$ from the following problem \vspace{-1mm}
\begin{equation}\label{eq:dual}
    \max_{\bm \xi}\min_{\bm \theta} \mathbb{E}_{{\bf x}\sim\mathcal{D}}[\mathcal{L}({\bf x};{\bm \theta},{\bm \xi})],
\end{equation}
where $\mathcal{L}({\bf x};{\bm \theta},{\bm \xi})$ is the Lagrangian function defined as \vspace{-1.5mm}
\begin{equation}\label{eq:Lar}
    \!\!\mathcal{L}({\bf x};{\bm \theta},{\bm \xi}) \!=\! F({\bf x},f_{\bf y}({\bf x};{\bm \theta})) +\! \sum_{i=1}^{N_C}\! f_{\lambda_i}\!({\bf x};{\bm \xi}) G_i({\bf x},f_{\bf y}({\bf x};{\bm \theta})),\!
\end{equation}
and $f_{\lambda_i}({\bf x};{\bm \xi})$ represents the $i$th element of the output $[\lambda_1,\cdots,\lambda_{N_C}]^T=f_{\bm \lambda}({\bf x};{\bm \xi})$.

After being well-trained, the resource allocation variable provided by $f_{\bf y}(\cdot\,; {\bm \theta})$ can achieve near-optimal objective value in \eqref{prob:whole} and satisfy the constraints with high probability \cite{EisenDual,SCJ}.


To improve OOD performance, we should introduce a cost function\footnote{In supervised learning, the cost function is the same as loss function, since the loss can indicate the performance of DNNs.} to find the worst-case distribution for DRO, which measures how worse the DNN performs on a distribution. In this paper, we focus on pursuing better objective on unseen distributions, and thus select  $F({\bf x},{\bf y})$ as the cost function.\footnote{If one seeks to further guarantee the constraints on unseen distributions, $F({\bf x},{\bf y})+\sum_{i=1}^{N_C}\max(0,G_i({\bf x},{\bf y}))$ can be used as the cost function, and the subsequent analysis is similar.}

Analogous to \eqref{eq:DRO}, one can optimize ${\bm \theta}$ on the distribution $\tilde{\mathcal{D}}$ with maximal cost, which is the solution of $\max_{\tilde{\mathcal{D}}\in \mathbb{D}} \mathbb{E}_{{\bf x}\sim \tilde{\mathcal{D}}} \left[ F({\bf x}, f_{\bf y}({\bf x}; {\bm \theta}))\right]$.
However, $\tilde{\mathcal{D}}$ obtained in this way is NOT the worst-case distribution, because this objective cannot differentiate the cost increase caused by \emph{the higher optimal objective value} and by \emph{the degraded generalization performance of} $f_{\bf y}(\cdot\,;{\bm \theta})$ on the distributions in $\mathbb{D}$.
The former is unrelated to generalizability.
For example, for SR-maximization resource allocation, a decrease in channel strength inevitably increases the cost (i.e., reduces SR), but a DNN may not perform worse on low-strength channels. Then, optimizing the DNN on such channels is not beneficial for OOD generalization.

To address this issue,\footnote{This issue does not arise in supervised learning, since the optimal value of the losses (such as cross-entropy, mean squared error) is always zero.} we reformulate DRO for PDL as
\vspace{-1mm}
\begin{subequations}\label{eq:newDRO}
\begin{align}
    &P_{\rm DRO}:~\max_{\bm \xi}\min_{\bm \theta} \mathbb{E}_{{\bf x}\sim\tilde{\mathcal{D}}}[\mathcal{L}({\bf x};{\bm \theta},{\bm \xi})]\label{eq:newDRO-a}\\
    {\rm s.t.}~&\tilde{\mathcal{D}} \!=\! \arg\max_{\mathcal{D}'\in \mathbb{D}}\! \mathbb{E}_{{\bf x}\sim \mathcal{D}'} \!\left[ F({\bf x}, f_{\bf y}({\bf x}; {\bm \theta})) \!-\! F({\bf x}, f_{{\bf y}^*}({\bf x}))\right]\!, \!\!\! \label{eq:newDRO-b}
\end{align}
\end{subequations}
for training $f_{\bf y}(\cdot\,;{\bm \theta})$ and $f_{\bm \lambda}(\cdot\,;{\bm \xi})$ on the \emph{truely worst distribution}, where the output of $f_{\bf y}(\cdot\,;{\bm \theta})$ yields the largest \emph{gap} in cost from the optimal solution instead of the highest cost.

Inspired by the proof in \cite{DRO-Adv-ICML2021}, we adopt AT as a realizable alternative to the DRO for enhancing OOD generalizability performance. By replacing $\tilde{\mathcal{D}}$ by $\tilde{\bf x}$, the AT problem for PDL is reformulated as
\vspace{-2mm}
\begin{subequations}\label{eq:newAT}
\begin{align}
    &P_{\rm AT}:~\max_{\bm \xi}\min_{\bm \theta} \mathbb{E}_{{\bf x}\sim\mathcal{D}}[\mathcal{L}(\tilde{\bf x};{\bm \theta},{\bm \xi})]\label{eq:newAT-L}\\
    {\rm s.t.}~&\tilde{\bf x}=\arg\!\!\!\!\max_{{\bf x}'\in B_{\epsilon}({\bf x})}\!\! \left[ F({\bf x}', f_{\bf y}({\bf x}'; {\bm \theta})) - F({\bf x}', f_{{\bf y}^*}({\bf x}'))\right].\! \label{eq:newAT-x}
\end{align}
\end{subequations}

\subsection{Unsupervised Adversarial Training (UAT) Method}
To solve $P_{\rm AT}$, i.e., to train $f_{\bf y}(\cdot\,; {\bm \theta})$ and $f_{\bm \lambda}(\cdot\,;{\bm \xi})$ with identified adversarial examples, \eqref{eq:newAT-x} and \eqref{eq:newAT-L} can be alternately optimized. To find accurate adversarial examples, prior to solving $P_{\rm AT}$, both DNNs need to first be trained with in-distribution (ID) samples, namely, original samples provided in the training dataset.

In what follows, we show how to find the adversarial example $\tilde{\bf x}$ for every original sample ${\bf x}$ from \eqref{eq:newAT-x}.

Since $\tilde{\bf x} = {\bf x} + {\bm \delta}$, we can find ${\bm \delta}$ instead.
From the definition of $B_{\epsilon}({\bf x})$ in \eqref{eq:AT}, \eqref{eq:newAT-x} can be transformed into the following problem to optimize ${\bm \delta}$
\begin{equation}
       \max_{\left\|{\bm \delta}\right\|_2\leq\epsilon}\! F({\bf x}+{\bm \delta}, f_{\bf y}({\bf x}+{\bm \delta}; {\bm \theta})) - F({\bf x}+{\bm \delta}, f_{{\bf y}^*}({\bf x}+{\bm \delta})).\!\! \label{eq:Taylor1}
\end{equation}

When $\epsilon$ is small, it can be seen that \eqref{eq:Taylor1} has the same solution as the following problem  by using the first-order Taylor polynomial,
\begin{equation}
       \max_{\left\|{\bm \delta}\right\|_2\leq\epsilon} \left(\frac{\partial F({\bf x}, f_{\bf y}({\bf x}; {\bm \theta}))}{\partial {\bf x}} - \frac{{\rm d}F({\bf x}, f_{{\bf y}^*}({\bf x}))}{{\rm d} {\bf x}}\right)^T{\bm \delta}. \label{eq:Taylor}
\end{equation}

The problem in \eqref{eq:Taylor} maximizes a linear objective over an Euclidean ball, whose optimal solution can be obtained as
\begin{equation}\label{eq:delta}
    {\bm \delta} = \epsilon \cdot \frac{\frac{\partial F({\bf x}, f_{\bf y}({\bf x}; {\bm \theta}))}{\partial {\bf x}} - \frac{{\rm d}F({\bf x}, f_{{\bf y}^*}({\bf x}))}{{\rm d} {\bf x}}}{\left\| \frac{\partial F({\bf x}, f_{\bf y}({\bf x}; {\bm \theta}))}{\partial {\bf x}} - \frac{{\rm d}F({\bf x}, f_{{\bf y}^*}({\bf x}))}{{\rm d} {\bf x}} \right\|_2}.
\end{equation}

To obtain the value of the numerator in \eqref{eq:delta} for given ${\bf x}$ and ${\bm \theta}$, we rewrite its first term by using the chain rule as
\begin{equation}\label{eq:part1}
    \frac{\partial F({\bf x}, f_{\bf y}({\bf x}; {\bm \theta}))}{\partial {\bf x}} = \frac{\partial F}{\partial {\bf x}} + \frac{\partial F}{\partial {\bf y}}\cdot \frac{\partial f_{\bf y}({\bf x};{\bm \theta})}{\partial {\bf x}},
\end{equation}
where the values of partial derivatives $\frac{\partial F}{\partial {\bf x}}$ and $\frac{\partial F}{\partial {\bf y}}$ can be computed using the value of ${\bf y}$ produced by $f_{\bf y}({\bf x};{\bm \theta})$, and $\frac{\partial f_{\bf y}({\bf x};{\bm \theta})}{\partial {\bf x}}$ can be obtained from the DNN due to its differentiability.

Obtaining the value of the second term of the numerator is challenging, due to the unknown function $f_{{\bf y}^*}(\cdot)$. To cope with this difficulty, we invoke the \emph{envelope theorem} \cite{EnvelopeTheorem}, which provides the gradient of the optimal objective from \eqref{prob:whole}, with no need for the expression of the optimal policy $f_{{\bf y}^*}(\cdot)$, as
\begin{equation}\label{eq:part2}
    \frac{{\rm d}F({\bf x}, f_{{\bf y}^*}({\bf x}))}{{\rm d} {\bf x}} = \frac{\partial F}{\partial {\bf x}}  + \lambda_{1}^* \frac{\partial G_{1}}{\partial {\bf x}} + \cdots + \lambda_{N_C}^* \frac{\partial G_{N_C}}{\partial {\bf x}},
\end{equation}
where $\lambda_{1}^*,\cdots,\lambda_{N_C}^*$ are the optimal Lagrange multipliers for the $N_C$ constraints in \eqref{prob:whole}.

Unlike \eqref{eq:part1}, the values of partial derivatives in \eqref{eq:part2} are computed with the value of optimal solution ${\bf y}^*$. In practice, the value of ${\bf y}^*$ can be approximated by ${\bf y}$ (i.e., the output of $f_{\bf y}({\bf x};{\bm \theta})$), since ${\bf x}$ herein is an ID sample and $f_{\bf y}(\cdot;{\bm \theta})$ has been trained with all ID samples before solving $P_{\rm AT}$.
Similarly, the values of $\lambda_{1}^*,\cdots,\lambda_{N_C}^*$ can be approximated by the outputs of multiplier network $f_{\bm \lambda}({\bf x};{\bm \xi})$.

By combining \eqref{eq:part1} and \eqref{eq:part2}, we have \vspace{-1mm}
\begin{equation}\label{eq:update-12}
    \!\!\!\frac{\partial F({\bf x}, f_{\bf y}({\bf x};\! {\bm \theta})\!)\!}{\partial {\bf x}} - \frac{{\rm d}F({\bf x}, f_{{\bf y}^*}\!({\bf x})\!)\!}{{\rm d} {\bf x}} \!=\! \frac{\partial F}{\partial {\bf y}}\cdot \frac{\partial f_{\bf y}}{\partial {\bf x}} - \!\sum_{i=1}^{N_C} \!f_{\lambda_{i}}\!\frac{\partial G_{i}}{\partial {\bf x}}.\!\!\!
\end{equation}
By substituting \eqref{eq:update-12} into \eqref{eq:delta}, the adversarial example of ${\bf x}$, i.e., $\tilde{\bf x} = {\bf x} + {\bm \delta}$, can be obtained as \vspace{-1mm}
\begin{equation}\label{eq:update}
    \!\!\tilde{\bf x} \!=\! {\bf x} \!+ \epsilon \cdot \frac{\frac{\partial F}{\partial {\bf y}}\!\cdot\! \frac{\partial f_{\bf y}({\bf x};{\bm \theta})}{\partial {\bf x}} \!-\!\! \sum_{i=1}^{N_C} \!f_{\lambda_{i}}\!({\bf x};{\bm \xi})\frac{\partial G_{i}}{\partial {\bf x}}}{\left\|\! \frac{\partial F}{\partial {\bf y}}\!\cdot\! \frac{\partial f_{\bf y}({\bf x};{\bm \theta})}{\partial {\bf x}} \!-\!\! \sum_{i=1}^{N_C} \!f_{\lambda_{i}}\!({\bf x};{\bm \xi})\frac{\partial G_{i}}{\partial {\bf x}} \!\right\|_2\!\!\!} \!\triangleq\! {\bf x} \!+ \epsilon \cdot d({\bf x}).\!\!
\end{equation}

\vspace{1mm} \emph{Remark 1:} The constraints in \eqref{prob:whole} for some wireless problems are imposed on ${\bf y}$ but not related to ${\bf x}$. For instance, power constraint is independent of channels. Such constraints can be ensured via projection methods, and render the multiplier network $f_{\bm \lambda}({\bf x};{\bm \xi})$ unnecessary. For these problems, $\frac{\partial G_i}{\partial {\bf x}}=0$ in \eqref{eq:part2}$\sim$\eqref{eq:update}. By contrast, for the problems with complex constraints (say the QoS constraints) that are related to ${\bf x}$ (say channels), $f_{\bm \lambda}({\bf x};{\bm \xi})$ is required and the values of corresponding partial derivatives $\frac{\partial G_i}{\partial {\bf x}}$ in \eqref{eq:update} ought to be computed. \vspace{1mm}

With a batch of adversarial examples, \eqref{eq:newAT-L} can be optimized by standard back-propagation algorithms (say Adam algorithm), where $\mathcal{L}(\cdot\,;{\bm \theta},{\bm \xi})$ is computed using the adversarial examples rather than the original samples.

During training, \eqref{eq:newAT-x} and \eqref{eq:newAT-L} are alternately solved for $T$ iterations. In the first iteration, the adversarial examples are identified based on the original samples using \eqref{eq:update}. In the subsequent iterations, the adversarial examples are identified based on those obtained in the previous iteration. In this way, the range of perturbations can be expanded for facilitating OOD generalization, while in each iteration $\epsilon$ can be set as a small value for the Taylor approximation to be accurate.


The proposed training method, referred to as UAT, is summarized in Algorithm~\ref{alg:AT}, where Lines 8$\sim$15 are the steps specifically for UAT. The lines marked with an asterisk (*) can be omitted if there are no constraints related to ${\bf x}$.

\emph{Remark 2:} The training complexity of UAT is $(T+1)$ times higher than PDL (i.e., Lines 1$\sim$7 in Algorithm~\ref{alg:AT}).


\begin{algorithm}
    \caption{Unsupervised Adversarial Training (UAT)}\label{alg:AT}
    \begin{algorithmic}[1]
        \Statex {\bf Input:} number of epochs $E$, number of batches $B$, batch-size $N_B$, samples $\{{\bf x}_n\}_{n=1}^{BN_B}$, initial value of weight ${\bm \theta}$, learning rate $\eta_1,\eta_2$, the epoch at which UAT begins $E_A$, number of adversarial iterations $T$, perturbation range $\epsilon$
        \Statex * initial value of weight ${\bm \xi}$ for multiplier network
        \For{$epoch = 1,\dots,E$}
            \For{$batch = 1,\dots,B$}
                \State $\mathcal{N}=\{1+N_B(batch-1),\cdots,N_B(batch)\}$
                \State $\mathcal{L} = \frac{1}{N_B}\sum_{n\in \mathcal{N}} F({\bf x}_n,f_{\bf y}({\bf x}_n; {\bm \theta}))$
                \State *$\mathcal{L} \!\leftarrow\! \mathcal{L} \!+ \frac{1}{N_B}\!\!\sum_{n}\!\! \sum_{i=1}^{N_C} \!f_{\lambda_i} ({\bf x}_n;{\bm \xi})G_i({\bf x}_n,f_{\bf y}({\bf x}_n; {\bm \theta}))$
                \State ${\bm \theta}\leftarrow {\bm \theta}-\eta_1 \frac{\partial \mathcal{L}}{\partial{\bm \theta}}$
                \State *${\bm \xi}\leftarrow {\bm \xi}+\eta_2 \frac{\partial \mathcal{L}}{\partial{\bm \xi}}$
                \If{$epoch\geq E_A$} \Comment{UAT begins}
                    \State $\tilde{\bf x}_{n}^{0} = {\bf x}_{n}, ~~\forall n\in \mathcal{N}$
                    \For{$t = 1,\cdots,T$}
                        \State\hspace{-4mm} $\tilde{\bf x}_{n}^{t} = \tilde{\bf x}_{n}^{t-1} \!+ \epsilon\cdot d(\tilde{\bf x}_{n}^{t-1}), \!\forall n \!\in\! \mathcal{N}$ \Comment{see \eqref{eq:update}}
                        \State\hspace{-4mm} $\mathcal{L} = \frac{1}{N_B}\sum_{n\in \mathcal{N}} F(\tilde{\bf x}_{n}^{t}, f_{\bf y}(\tilde{\bf x}_{n}^{t}; {\bm \theta}))$
                        \State\hspace{-4mm} *$\!\mathcal{L} \!\! \leftarrow \!\! \mathcal{L} \!+\! \frac{1}{N_B}\!\!\sum_{n}\!\! \sum_{i=\!1}^{\!N_C}\!\! f_{\!\lambda_i}\!(\!\tilde{\bf x}_n^t;\!{\bm \xi}\!) G_i\!(\tilde{\bf x}_n^t,\!f_{\bf y}(\tilde{\bf x}_{n}^{t}; \!{\bm \theta})\!)$
                        \State\hspace{-4mm} ${\bm \theta}\leftarrow {\bm \theta}- \eta_1 \frac{\partial \mathcal{L}} {\partial{\bm \theta}}$ \Comment{via back-propagation}
                        \State\hspace{-4mm} *${\bm \xi}\leftarrow {\bm \xi}+ \eta_2 \frac{\partial \mathcal{L}}{\partial{\bm \xi}}$ \Comment{via back-propagation}
                    \EndFor
                \EndIf
            \EndFor
        \EndFor
        \vspace{-1mm}
        \Statex {\bf Output:} trained weight ${\bm \theta}$
    \end{algorithmic}
\end{algorithm}

\vspace{-2mm}
\section{Case Study: Optimizing Hybrid Precoding} \label{sec:case}\vspace{-0mm}
In this section, we apply UAT for optimizing QoS-constrained hybrid precoding.

Consider a base station (BS) equipped with $N_T$ antennas and $N_{RF}$ radio frequency (RF) chains, which serves $K$ single-antenna users. The analog and baseband precoders can be jointly optimized to maximize SR as follows \cite{HYqos}, \vspace{-1mm}
\begin{subequations}\label{prob:HBF}
	\begin{align}
	&\max_{{\bf W}_{RF},{\bf W}_{BB}} ~~ \textstyle\sum_{k=1}^K R_k  \label{prob:HBF-obj}\\
	&{\rm s.t.}~~\left\|{\bf W}_{RF} {\bf W}_{BB} \right\|^2_F = P_{tot},\label{prob:HBF-power}\\
	&~~|\left({\bf W}_{RF}\right)_{n,j}|=1,n=1,\cdots,N_T,j=1,\cdots,N_{RF},\label{prob:HBF-modulus}\\
        &~~R_k\geq \gamma_k,k=1,\cdots,K,\label{prob:HBF-qos}
	\end{align}
\end{subequations}
where $R_k\!=\!\log_2\!\left(\!1\!+\!\frac{|{\bf h}^H_k {\bf W}_{RF} {\bf w}_{BB_k}|^2}{\sum_{i=1,i\neq k}^K\! |{\bf h}^H_k {\bf W}_{RF} {\bf w}_{BB_i}|^2\!+\sigma^2} \!\!\right)$, ${\bf h}_k \in \mathbb{C}^{N_T}$, and $\gamma_k$ are the achievable rate,  channel vector, and data-rate requirement of the $k$th user, respectively, ${\bf W}_{RF} \in \mathbb{C}^{N_T\times N_{RF}}$ is the analog precoder, ${\bf W}_{BB}=[{\bf w}_{BB_1},\cdots,{\bf w}_{BB_K}]\in \mathbb{C}^{N_{RF} \times K}$ is the baseband precoder, $P_{tot}$ is the total power, and $\sigma^2$ is the noise power. \eqref{prob:HBF-power} is the power constraint, \eqref{prob:HBF-modulus} is the constant modulus constraint for the analog precoder, and \eqref{prob:HBF-qos} is the QoS constraint for every user.

In this task, ${\bf H}=[{\bf h}_1,\cdots{\bf h}_K]^T\in \mathbb{C}^{K\times N_T}$ is the environment parameter. To apply a DNN to the systems with different transmit power budgets and noise power levels, $S\!N\!R \triangleq ||{\bf H}||_F^2 P_{tot}/\sigma^2$ is also regarded as an input. Then, the DNN for learning the hybrid precoding is $({\bf W}_{RF},{\bf W}_{BB})=f_{\bf y}({\bf H}/\!\left\|{\bf H}\right\|_F, S\!N\!R;{\bm \theta})$ with two inputs, where the channel is normalized to avoid redundancy in the input information. Meanwhile, since both \eqref{prob:HBF-power} and \eqref{prob:HBF-modulus} are unrelated to ${\bf H}$ and can be guaranteed by the projection method in \cite{LSJ_MDGNN}, only the $K$ QoS constraints in \eqref{prob:HBF-qos} need to be satisfied using a multiplier network denoted by $[\lambda_1,\cdots,\lambda_K]^T =f_{\bm \lambda}({\bf H}/\!\left\|{\bf H}\right\|_F, S\!N\!R;{\bm \xi})$.

When applying UAT, we only need to identify adversarial examples for ${\bf x}={\bf H}/\!\left\|{\bf H}\right\|_F$, since $S\!N\!R$ is a scalar with finite range that does not need to be generalized. To ensure that the adversarial examples are normalized, we modify \eqref{eq:update} as
\begin{equation}
    \tilde{\bf x}=\frac{{\bf x} + \epsilon \cdot d({\bf x})}{\left\|{\bf x} + \epsilon \cdot d({\bf x})\right\|_2}.
\end{equation}

\section{Simulation Results}\label{sec:result}
In this section, we evaluate the OOD performance of DNNs trained by the proposed UAT for the hybrid precoding task.

We use the following two metrics to measure performance. Average SR (ASR) is the averaged SR over a test set, where the SR of a sample is recorded as zero if one or more constraints are not satisfied. Violation Ratio (VR) is the proportion of samples in a test set that violate the constraints.

To obtain a benchmark for comparison, we simulate a difference-of-convex (DC) programming-based numerical algorithm \cite{HYqos} to optimize hybrid precoders. Yet different from  \cite{HYqos}, we set ${\bf W}_{RF}$ as the phases of the first $N_{RF}$ right singular vectors of ${\bf H}$, instead of selecting from the array responses computed by the angles of departure (AoDs) of all propagation paths that is only applicable to sparse channels.

In simulations, the BS equipped with $N_T=16$ antennas and $N_{RF}=6$ RF chains serves $K=4$ users with data-rate requirements of $\gamma_1=\cdots=\gamma_K=1$ bps/Hz. $P_{tot}/\sigma^2$ is uniformly selected from the range of 0$\sim$20 dB during training, and is set as 10 dB for testing unless otherwise specified.

We compare UAT with two existing AT methods in \cite{PGD,DRO-Adv-ICML2021} that generate adversarial examples for supervised learning.
Following the configurations in \cite{PGD,DRO-Adv-ICML2021}, the number of descent steps is 7, step size is 2/64, projection radius is 8/64. 
For UAT, the fine-tuned hyper-parameters are $E_A\!=\!50$, $T\!=\!5$, $\epsilon\!=\!5/64$. The PDL \cite{SCJ} without AT serves as a baseline.

Both $f_{\bf y}(\cdot\,;{\bm \theta})\!$ and $f_{\bm \lambda}(\cdot\,;{\bm \xi})\!$ are realized by three kinds of DNNs designed for hybrid precoding: convolutional neural network (CNN) \cite{CNNprecode}, three-dimensional graph neural network (3D-GNN) \cite{LSJ_MDGNN}, and recursive GNN (RGNN) \cite{RGNN}. The fine-tuned hyper-parameters are as follows.
The optimizer is Adam, $\eta_1=\eta_2=0.001$,  $N_B=100$, batch normalization is used for CNN and 3D-GNN, activation function is ${\rm Tanh}(\cdot)$ for RGNN and ${\rm ReLU}(\cdot)$ for CNN and 3D-GNN. Other fine-tuned hyper-parameters are listed in Table~\ref{table:NN}, where $3\times3$ convolution kernels are used in the hidden layers of CNN followed by a fully-connected layer with 512 neurons.

\begin{table}[htb!]
\vspace{-0mm}
\setlength\tabcolsep{0.12pt}
\centering
\caption{Fine-Tuned Hyper-Parameters for DNNs}\label{table:NN}
\vspace{-2.5mm}
\footnotesize
    \begin{tabular}{c|c|c|c|c}
        \hline\hline
            \multirow{2}{*}{\makecell{\bf Kind \\ \bf of \\ \bf DNN}} & \multicolumn{2}{c|}{$f_{\bf y}(\cdot~;{\bm \theta})$} & \multicolumn{2}{c}{$f_{\bm \lambda}(\cdot~;{\bm \xi})$} \\ \cline{2-5}
           & \makecell{\bf Number of \\\bf hidden layers} & \makecell{\bf Number of \\ \bf channels in layers} & \makecell{\bf Number of \\ \bf hidden layers} & \makecell{\bf Number of \\ \bf channels in layers} \\ \hline
          CNN & 5 & [256, $\cdots$, 256, 16] & 3 & [128, 128, 16] \\ \hline
          3D-GNN & 6 & [256, $\cdots$, 256] & 4 & [128, $\cdots$, 128] \\ \hline
          RGNN & 4 & [64, 64, 64, 16] & 3 & [32, 32, 8]  \\ \hline
          \hline
    \end{tabular}
\vspace{-6mm}
\end{table}

Since the UAT does not depend on data from different distributions, these DNNs are \emph{trained} with 100,000 samples generated from a single distribution, for which we take Rayleigh fading channel as an example. In fact, they can also be trained on other channel distributions and UAT remains effective. \emph{Once trained, these DNNs are not fine-tuned}, in order to examine their OOD generalization performance.

The performances of the DNNs trained with different methods are evaluated on several test sets, each consisting of 2,000 samples generated from unseen channel distributions (referred to as OOD test samples), as detailed in the sequel.

\subsection{Rician Fading Channel}
The OOD test samples are generated from the Rician fading channel model:
${\bf h}_k = \sqrt{{\kappa}/({\kappa+1})}{\bf a}(\phi_k)+\sqrt{{1}/({\kappa+1})}{\bf z}_k$,
where ${\bf a}(\phi_k)=[1,{\rm e}^{{\rm i}\pi\sin(\phi_k)},\cdots,{\rm e}^{{\rm i}(N_T-1)\pi\sin(\phi_k)}]^T$ is the array response of the line-of-sight (LOS) path with AoD $\phi_k\sim \mathcal{U}(0,2\pi)$, ${\bf z}_k\sim \mathcal{CN}({\bf 0},{\bf I}_{N_T})$, and $\kappa$ is the Rician factor.

\begin{figure}[!htb]
\vspace{-0mm}
    \centering
    \subfloat[ASR performance]{
        \begin{minipage}[c]{0.45\linewidth}
            \centering
            \includegraphics[width=\linewidth]{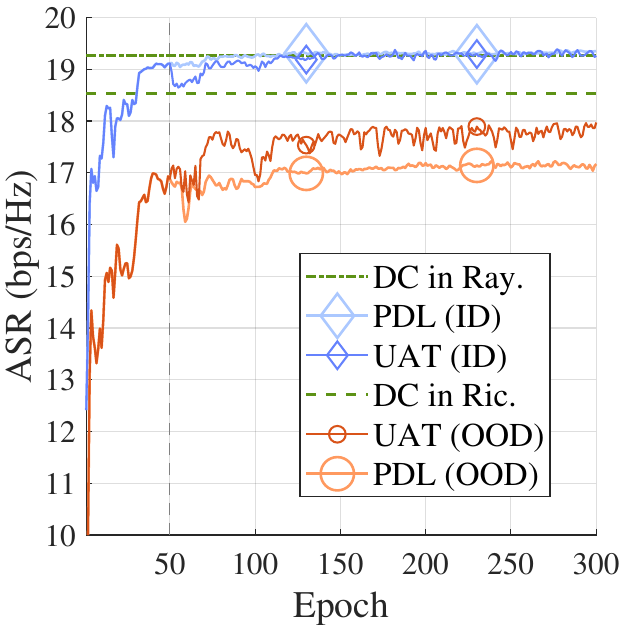}
            \label{fig:epoch-a}
            \vspace{-5.5mm}
        \end{minipage}
    }
    \subfloat[VR performance]{
        \begin{minipage}[c]{0.45\linewidth}
            \label{fig:epoch-b}
            \centering
            \includegraphics[width=\linewidth]{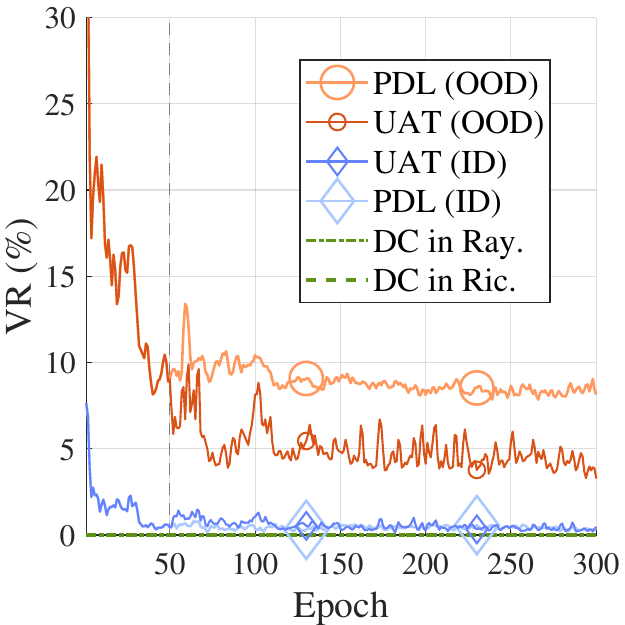}
            \vspace{-5.5mm}
        \end{minipage}
    }
    \vspace{-1mm}
    \caption{Performance of RGNN on ID test samples (Rayleigh) and OOD test samples (Rician, $\kappa=10 $ dB), during training with PDL or UAT.}
    \label{fig:epoch}
\vspace{-5mm}
\end{figure}

In Fig.~\ref{fig:epoch}, we provide the performance of the RGNN trained by PDL or UAT (the results for the other two DNNs are similar and hence are not shown). It has legend PDL(ID) or UAT(ID) when tested in Rayleigh channel and it has legend PDL(OOD) or UAT(OOD) when tested in Rician channel. The Rician channel with $\kappa=10 $ dB is a representative small-scale fading of LOS channels in UMa and UMi scenarios \cite{3gpp38901}. Fig.~\ref{fig:epoch}~(a) shows that RGNN trained by both methods can achieve comparable ASR to DC in Rayleigh channel. When tested in Rician channel, the RGNN trained by UAT attains a 4.7\% higher ASR (normalized by DC) than that trained by PDL.
Fig.~\ref{fig:epoch}~(b) shows that DC can satisfy QoS constraints for all samples on both channels. When tested in Rayleigh channel, RGNN trained by both methods exhibits extremely low VR  (about 0.3\%). When tested in Rician channels, the RGNN trained by UAT has a 4.9\% lower VR than that trained by PDL, because the objective in \eqref{prob:HBF-obj} and the constraints in \eqref{prob:HBF-qos} are positively correlated, even though no constraint satisfaction is considered in the cost of our DRO formulation.

Besides, the inference time of RGNN is 15.48 ms, but DC requires 5498 ms on the same CPU, validating the advantage of learning-based approach for low decision latency.

\begin{figure}[!htb]
\vspace{-0mm}
    \centering
    \subfloat[ASR performance]{
        \begin{minipage}[c]{0.45\linewidth}
            \label{fig:ric-a}
            \centering
            \includegraphics[width=\linewidth]{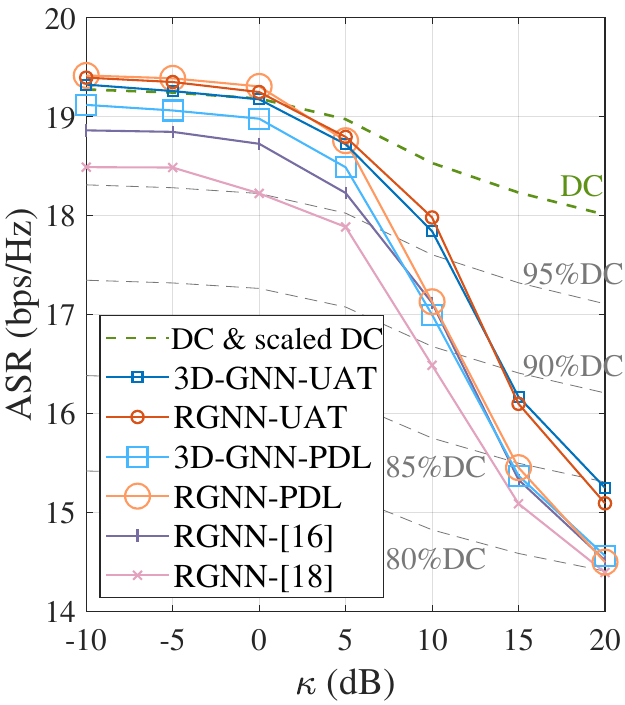}
            \vspace{-5.5mm}
        \end{minipage}
    }
    \subfloat[VR performance]{
        \begin{minipage}[c]{0.45\linewidth}
            \label{fig:ric-b}
            \centering
            \includegraphics[width=\linewidth]{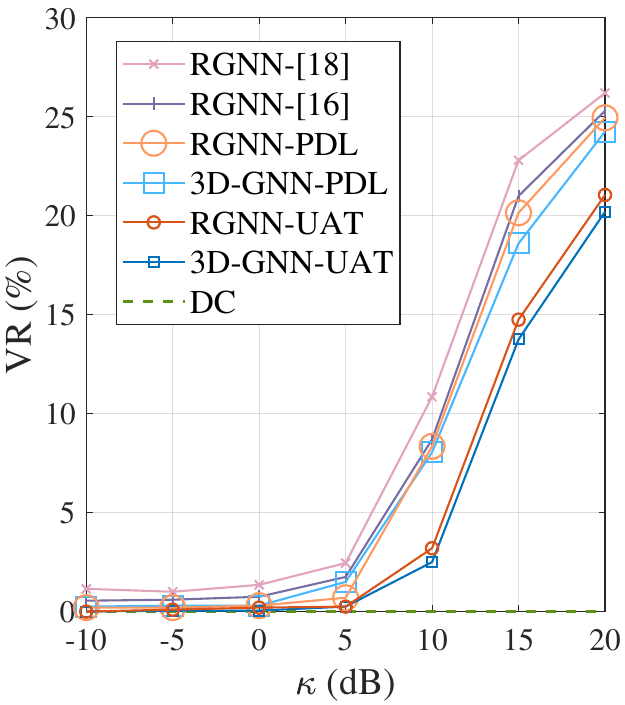}
            \vspace{-5.5mm}
        \end{minipage}
    }
    \vspace{-1mm}
    \caption{OOD performance of two GNNs trained with different methods.}
    \label{fig:ric}
\vspace{-5mm}
\end{figure}

In Fig.~\ref{fig:ric}, we compare the OOD generalization performance of RGNN and 3D-GNN trained by different methods. We can see that UAT improves OOD generalization by approximately 5\% in both ASR and VR for the two GNNs when the distribution of test samples differs noticeably from Rayleigh channel (say $\kappa\geq10$ dB).
The ASR of UAT-trained DNNs still retains 85\% of the ASR of DC even when $\kappa=20$~dB.
The DNNs trained by the AT methods in \cite{PGD,DRO-Adv-ICML2021} exhibit marginal OOD generalization gains over PDL, due to not designed for unsupervised learning.
Their performance degradation observed at $\kappa \leq 5$~dB (close to Rayleigh channels, i.e., ID samples) results from their overly broad perturbation range.

\subsection{Spatially Correlated Channel}
The OOD test samples are generated from the following spatially correlated channel model \cite{corrChannel}:
${\bf H} = {\bf R}^{\frac 1 2}_{U}{\bf Z}{\bf R}^{\frac 1 2}_{A}$,
where each element in ${\bf Z}$ is $z_{k,n}\sim \mathcal{CN}(0,1)$, ${\bf R}_{A}\in\mathbb{R}^{N_T\times N_T}$ is the antenna correlation matrix with element in the $i$th row and $j$th column being $\rho_A^{|i-j|}$, ${\bf R}_{U}\in\mathbb{R}^{K\times K}$ captures the inter-user correlation with diagonal elements being 1 and off-diagonal elements being $\rho_U$, and $\rho_A,\rho_U\in[0,1]$.

\begin{table}[htb!]
\vspace{-1mm}
\setlength\tabcolsep{1pt}
\centering
\caption{ASR Performance (Relative to DC) in Correlated Channels}\label{table:cor}
\vspace{-2.5mm}
\footnotesize
    \begin{tabular}{c|c|c|c|c|c|c|c|c|c}
        \hline\hline
        \multicolumn{2}{c|}{\multirow{2}{*}{}} & \multicolumn{4}{c|}{$\rho_U$ varies ($\rho_A=0$)} & \multicolumn{4}{c}{$\rho_A$ varies ($\rho_U=0$)} \\ \cline{3-10}
        \multicolumn{2}{c|}{} & 0.2 & 0.4 & 0.6 & 0.8 & 0.2 & 0.4 & 0.6 & 0.8 \\ \hline
        \multicolumn{2}{c|}{DC (bps/Hz)} & 18.77 & 17.62 & 15.65 & 12.36 & 19.13 & 18.86 & 18.14 & 16.26 \\ \hline
        \multirow{2}{*}{RGNN} & PDL & 100.3\% & 99.2\% & 91.3\% & 72.6\% & 100.6\% & 99.1\% & \textcolor{Red}{95.6\%} & \textcolor{ForestGreen}{78.3\%} \\ \cline{2-10}
        ~ & UAT & 99.9\% & 98.8\% & \bf 95.8\% & \bf 87.4\% & 100.4\% & 100.3\% & \textcolor{Red}{\bf 99.7\%} & \textcolor{ForestGreen}{\bf 98.6\%} \\ \hline
        \multirow{2}{*}{3D-GNN} & PDL & 98.7\% & 97.3\% & 93.8\% & \textcolor{ForestGreen}{79.2\%} & 99.1\% & 98.5\% & \textcolor{Red}{96.4\%} & 84.9\% \\ \cline{2-10}
        ~ & UAT & 100.0\% & 99.4\% & \bf 98.1\% & \textcolor{ForestGreen}{\bf 93.0}\% & 100.3\% & 99.9\% & \bf \textcolor{Red}{99.5\%} & \bf 95.5\% \\ \hline
        \multirow{2}{*}{CNN} & PDL & 82.2\% & 80.0\% & 74.2\% & \textcolor{ForestGreen}{57.6\%} & \textcolor{Red}{82.6\%} & 81.9\% & 80.2\% & 70.5\% \\ \cline{2-10}
        ~ & UAT & 92.5\% & 91.3\% & 89.2\% & \textcolor{ForestGreen}{80.2\%} & \textcolor{Red}{92.8\%} & 92.1\% & 91.3\% & 84.0\% \\ \hline
        \hline
    \end{tabular}
\vspace{-2mm}
\end{table}

In Table~\ref{table:cor}, we provide the OOD generalization performance of the DNNs trained by PDL or UAT when they are tested on correlated channels. As shown by the values in boldface, the OOD generalization gains of UAT over PDL are more pronounced in highly correlated channels ($\rho_U,\rho_A\geq0.6$), which often arise in scenarios with densely distributed users, limited scatters, or closely spaced antennas.
In particular, for RGNN, the OOD gain ranges from 4.1\% (i.e., $99.7\%-95.6\%$, shown in red) to 20.3\% (shown in green, similarly hereafter), whereas for 3D-GNN, the gain ranges from 3.1\% to 13.8\%. The lower gain of 3D-GNN is attributed to its inherent relatively better OOD generalizability.
For CNN, UAT significantly boosts ASR by 10.2\%$\sim$22.6\% across all values of $\rho_U$ and $\rho_A$, not limited to highly correlated channels. This is because UAT can mitigate the overfitting of CNN.

\subsection{Sparse Channel and 3GPP Channel}
Finally, the OOD test samples are generated from the following two millimeter-wave channel models, which are commonly used for hybrid precoding.

The first channel model is an $L$-path sparse channel:
 $   {\bf h}_k = \sqrt{N_T/L}\textstyle\sum_{l=1}^L\alpha_{k,l}{\bf a}(\phi_{k,l})$,
where $\alpha_{k,l}\sim \mathcal{CN}(0,1)$ is a complex gain, and ${\bf a}(\phi_{k,l})$ is the array response for AoD $\phi_{k,l}\sim\mathcal{U}(0,2\pi)$. The second channel model is specified in 3GPP TR 38.901 \cite{3gpp38901} for UMa and UMi scenarios with a carrier frequency of 28 GHz.\footnote{$P_{tot}$ is 49 dBm for UMa and 44 dBm for UMi scenarios. In LOS scenarios, $P_{tot}$ is further reduced by 20 dB to avoid the extra high received power that results in nonlinear distortion. $\sigma^2$ is -85 dBm. Owing to the difference in large-scale fading channels among users, QoS constraints are hard to be satisfied and thus are omitted in these scenarios.}

\begin{table}[htb!]
\vspace{-1mm}
\setlength\tabcolsep{0.8pt}
\centering
\caption{ASR Performance (Relative to DC) in Sparse \& 3GPP Channels}\label{table:sparse}
\vspace{-2.5mm}
\footnotesize
    \begin{tabular}{c|c|c|c|c|c|c|c|c|c}
        \hline\hline
        \multicolumn{2}{c|}{\multirow{3}{*}{}} & \multicolumn{4}{c|}{Sparse Channels} & \multicolumn{4}{c}{Scenarios in TR 38.901} \\ \cline{7-10}
        \multicolumn{2}{c|}{} & \multicolumn{4}{c|}{Number of paths, $L$} & \multirow{2}{*}{\makecell{UMa\\NLOS}}&  \multirow{2}{*}{\makecell{UMa\\LOS}} &  \multirow{2}{*}{\makecell{UMi\\NLOS}} &  \multirow{2}{*}{\makecell{UMi\\LOS}} \\ \cline{3-6}
        \multicolumn{2}{c|}{} & 5 & 4 & 3 & 2 & & & & \\ \hline
        \multicolumn{2}{c|}{DC (bps/Hz)} & 18.28 & 18.24 & 17.93 & 17.43 & 9.74 & 15.44 & 14.89 & 16.84 \\ \hline
        \multirow{2}{*}{RGNN} & PDL & 96.6\% & 95.0\% & 89.7\% & 80.2\% & 95.0\% & 88.1\% & 73.8\% & 82.6\% \\ \cline{2-10}
        ~ & UAT & 97.8\% & 96.1\% & 92.7\% & 84.6\% & 111.3\% & 100.5\% & 91.7\% & 96.1\% \\ \hline
        \multirow{2}{*}{3D-GNN} & PDL & 96.4\% & 95.7\% & 93.1\% & 86.0\% & 97.0\% & 91.5\% & 76.4\% & 87.2\% \\ \cline{2-10}
        ~ & UAT & \bf 99.1\% & \bf 98.4\% & \bf 96.6\% & \bf 90.2\% & \bf 117.2\% & \bf 103.9\% & \bf 95.1\% & \bf 100.3\% \\ \hline
        \multirow{2}{*}{CNN} & PDL & 77.4\% & 75.8\% & 70.5\% & 60.8\% & 95.9\% & 77.8\% & 69.4\% & 74.2\% \\ \cline{2-10}
        ~ & UAT & 87.9\% & 86.0\% & 80.8\% & 70.3\% & 102.6\% & 86.2\% & 77.3\% & 82.7\% \\ \hline
        \hline
    \end{tabular}
\vspace{-2mm}
\end{table}

In Table~\ref{table:sparse}, we provide the OOD generalization performance of the DNNs trained by PDL or UAT when they are tested on the two kinds of channels. It is evident that UAT enhances the OOD generalizability of all three DNNs, under all considered channels for testing. The results in boldface indicate that 3D-GNN outperforms both RGNN and CNN in terms of OOD generalization.

\section{Conclusions}\label{sec:con}
In this paper, we reformulated the AT problem for training OOD-generalizable DNNs in the unsupervised manner when optimizing wireless resource allocation, and proposed a UAT method to solve the problem.
To validate the effectiveness of UAT, we use it to train three representative DNNs for learning QoS-aware hybrid precoding as a case study.
Simulation results showed that these DNNs trained by the proposed UAT on Rayleigh fading channels can achieve superior OOD performance on Rician fading, spatially correlated, sparse, and 3GPP channels. In future works, we will evaluate the OOD generalization performance over more realistic channels, such as real-world datasets.

\bibliography{IEEEabrv,ref}

\end{document}